\documentclass[prb,twocolumn,showpacs,amsmath,amssymb]{revtex4}

\usepackage{amsmath}
\usepackage{graphicx}
\usepackage[pdftex,dvips]{epsfig,graphicx}

\newcommand{\be}{\begin{equation}}
\newcommand{\ee}{\end{equation}}
\newcommand{\bma}{\begin{displaymath}}
\newcommand{\ema}{\end{displaymath}}
 
\begin{document}

\title{Magnetism of quantum dot clusters: A Hubbard model study}

\author{J.-P. Nikkarila$^{1,2}$, M. Koskinen$^1$ and M. Manninen$^1$}

\affiliation{\sl $^1$NanoScience Center, Department of Physics,
FIN-40014 University of Jyv\"askyl\"a, Finland}

\affiliation{\sl $^2$Inspecta LTD, FI-02151 Espoo, Finland}

 
\date{\today}

\begin{abstract} 
Magnetic properties of two and three-dimensional clusters
of quantum dots are studied with exact diagonalization of 
a generalized Hubbard model. 
We study the weak coupling limit, where the electrons
interact only within a quantum dot and consider cases
where the second or third harmonic oscillator shell 
is partially filled. The results show that in the case of half-filled shell 
the magnetism is determined by the antiferromagnetic 
Heisenberg model with spin 1/2, 1 or 3/2, depending on
the number of electrons in the open shell. 
For other fillings the system in most cases favors 
a large total spin, indicating a ferromagnetic coupling
between the dots.

\end{abstract}

\maketitle

\section{introduction}

Experiments have shown that cluster structure can show
stronger form of magnetism than the bulk structure of the same 
element. This interesting behavior has been 
reported for example for Fe, Ni and Co
clusters \cite{deHeer1990,Billas1994,Bucher1991,Douglass1993,Apsel1996}.
The size of clusters in these experiments has
varied from few to few hundreds of atoms.
Also Ga has been shown to exhibit paramagnetic behavior with many
cluster sizes \cite{Douglass1993}.
The experiments have started an intensive theoretical 
epoch in material physics, where for example the Ising model\cite{merikoski1991},
the Hubbard model \cite{pastor1994,lopezurias1999},
and different forms of the Density Functional Theory (DFT),
are applied \cite{koskinen1997,reimann2002,Pastor2003,Pastor2004}.

Experimental breakthroughs have also been achieved
in constructing artificial lattices from quantum dots
\cite{lee2001,schmidbauer2006,kohmoto2002}.
The artificial quantum dot molecules
and normal molecules (or clusters) have many differences,
where the most important one is that the geometrical
structure of an artificial molecule is fixed and the
degeneracy can not be removed by Jahn-Teller deformation.
This gives room for internal symmetry breaking through
spontaneous magnetism,
superconductivity or superfluidity.
Recent experiments have inspired much theoretical
work\cite{chen1997,taut2000,yannouleas2002,gu2007,xianlong2007,koponen2006,koponen2007,massel2005,paananen
2008a,chin2006,rom2006,nikkarila2008a}.

For a lattice with strongly correlated particles 
the generic many-particle model is
the Hubbard model, which has been vastly studied in the case of one state
per lattice site (for reviews see\cite{voit1994,kolomeisky1996}).
In a one-dimensional lattice or in a ring the Hubbard model is exactly solvable using the
Bethe ansatz\cite{lieb1968}.
The magnetism of finite molecules\cite{pastor1994,lopezurias1999}
and quantum rings\cite{yu1992,viefers2004} have also been studied using 
the simple Hubbard model.

Mean-field calculations based on the spin-density functional
theory, as well as ab initio configuration interaction calculations, 
predict that the Hund's first
rule determines the total spin of an isolated quantum dot,
independent of the interparticle interaction and confining
potential of the lattice site\cite{koskinen1997,manninen2001,reimann2002}. 
The magnetism of the 
quantum dot molecule then depends on the total spin of the individual
dots, on the geometry of the molecule and 
on the coupling between the quantum 
dots\cite{kolehmainen2000,koskinen2003,karkkainen2007,karkkainen2007a,karkkainen2007b}.
A simple tight-binding model (with the exchange splitting as a parameter) explains
qualitatively some of the magnetic properties of quantum dot lattices\cite{koskinen2003b}.

Single quantum dots with a few electrons can be solved numerically exactly
(i.e. to a high degree of convergence with respect to the 
necessary restrictions
in Hilbert space) by diagonalizing the many-body Hamiltonian 
(for a review see Ref.\cite{reimann2002}).
Methods beyond the mean-field approximation have also been applied to quantum dot
molecules\cite{yannouleas1999,bayer2001,harju2002,mireles2006,scheibner2007,zhang2007}.

The purpose of this paper is to study the magnetism of
quantum dot molecules with a generalized Hubbard model.
We consider dot molecules with two to four dots in two and
three-dimensional geometries and up to 12 electrons per dot.  
We assume the confining potential in each dot to be 
two-dimensional or three dimensional
and harmonic at the bottom. In this case the electrons 
in one dot fill the harmonic oscillator shells. We consider the
three lowest shells $1s$, $1p$, and $2s1d$.
We use a generalization of the simple Hubbard model to describe the
interactions: The particles interact only within a quantum dot and
only the partially filled shell is included in the Hilbert space.
The Hubbard Hamiltonian is then solved with exact diagonalization.

\section{Theoretical models}

\subsection{General Hubbard model}

We assume a generalized Hubbard model Hamiltonian
\be
\hat{H}=\hat{J}+\hat{U},
\label{hamiltonian}
\ee
where the first term represents hopping between neighboring 
sites and the second term intra-site two-body interactions. 

Hoppings preserve spin, and are equal for spin-up and spin-down 
particles. Thus, $\hat J$ separates into two symmetric spin parts:
$\hat J = \sum _{\sigma = \uparrow, \downarrow} \hat J _{\sigma }$.
The hopping part of the Hamiltonian is generally of the form
\begin{equation}
\hat {J}_\sigma = - \sum _{nn'} \sum _{jj'} J_{nn'jj'} 
c^{\dagger } _{nj\sigma } c _{n'j' \sigma } ,
\label{hopping}
\end{equation}
where $n$ and $n'$ refer to different dots, whereas $j$ and $j'$ denote the 
orbitals in question, in $n$ and $n'$, respectively.
In the case of $p$ and $d$ orbitals we consider directional dependence of the 
hopping integrals $J$ (see below).
In the case of a square shaped four-dot molecule we assume that
there is no diagonal hopping.

The two-body interactions of the Hubbard model are treated 
in the spirit of the tight-binding model:
The electrons only interact when they are at the same quantum dot.
Thus, $\hat U$ separates in the symmetric parts representing interactions on
each site $n$: $\hat U = \sum _n \hat U _n$.
Within a site, full (spin-independent) two-body interaction is considered, 
which yields
\begin{equation}
\hat U _n = {1\over 2} \sum _{\stackrel{j_1j_2j_3j_4}{\sigma \sigma '}}
U_{j_1j_2j_3j_4} 
c_{nj_1\sigma }^{\dagger }
 c_{nj_2\sigma ' }^{\dagger }
 c_{nj_4\sigma ' }
 c_{nj_3\sigma  }
\label{interact} 
\end{equation}
where $U_{j_1j_2j_3j_4}$ are the direct space matrix elements 
of on-site interaction, depending on the interaction itself
and the $j$-orbitals in question, i.e. the eigenstates of the confining 
potential. 

We do not take advantage of the fact that the Hamiltonian
does not depend on spin, but diagonalize the system for $S_z=0$
for even number of electrons and for $S_z=1/2$ for odd number
of electrons and only afterwards determine the total spin $S$ for each 
many-particle state. The total number of electrons is denoted by $N$.
Lanczos method was used to diagonalize the matrix.
The largest matrix dimensions, for four dot systems with 12 electrons,
were more than 800000 with more than 1.5 billion nonzero elements.

\subsection{On-site interactions}

We consider mainly two-dimensional systems and
two kinds of interactions between the electrons,
the delta function interaction and the Coulomb interaction.
In the case of the delta function interaction and harmonic confinement,
the relative ratios of the
matrix elements $U_{j_1j_2j_3j_4}$ (Eq. (\ref{interact})), for each electron shell $k$, 
can be characterized by just one constant $U_k$.
Furthermore, for the $p$-shell the ratio of the two different
matrix elements is independent of the radial form of the 
confining potential.
In the case of the Coulomb interaction we have determined the 
interaction matrix elements only for the harmonic confinement.
We consider only the open shell in each dot to be in the active 
Hilbert space and only need the interaction matrix elements in
each shell separately. These are given in Table \ref{umatrix}.
Notice that for the $p$-shell the only difference in going from the
short-range delta function interaction to the long-range Coulomb interaction
is the reduction of one $U$ value as compared to the others.

\begin{table}
\centering
\linespread{1}
\small
\caption{ 
Nonzero interaction matrix elements for each energy shells of a two-dimensional
harmonic confinement calculated with the delta function interaction ($U^\delta$)
and with the Coulomb interaction ($U^C$). $U_k$ determines the 
strength of the contact interaction. In the case of the Coulomb interaction 
the matrix elements are given in atomic units for harmonic potential with
eigenfrequency $\omega_0=1$ atomic units. For $p$-electrons we show for the 
delta function interaction the matrix elements for two representations: For $p_x$ and $p_y$
and for the angular momentum states $p_-$ and $p_+$.
}
\label{umatrix}
\begin{tabular}{ccccccc}
shell& $j_1$ & $j_2$ & $j_3$ & $j_4$ & $U_{j_1j_2j_3j_4}^\delta$ & $U_{j_1j_2j_3j_4}^C$\\
\hline
1& $s$ & $s$ & $s$ & $s$ & $U_1$ &  1.2533\\
\hline
2&$p_x$ & $p_x$ & $p_x$ & $p_x$ & $3U_2$ & \\
2&$p_x$ & $p_x$ & $p_y$ & $p_y$ & $U_2$ & \\
2&$p_x$ & $p_y$ & $p_x$ & $p_y$ & $U_2$ & \\
2&$p_x$ & $p_y$ & $p_y$ & $p_x$ & $U_2$ & \\
2&$p_y$ & $p_y$ & $p_y$ & $p_y$ & $3U_2$ & \\
\hline
2&$p_-$ & $p_-$ & $p_-$ & $p_-$ & $2U_2$ & 0.86160\\
2&$p_-$ & $p_+$ & $p_-$ & $p_+$ & $2U_2$ & 0.86160 \\
2&$p_-$ & $p_+$ & $p_+$ & $p_-$ & $2U_2$ & 0.23494 \\
2&$p_+$ & $p_+$ & $p_+$ & $p_+$ & $2U_2$ & 0.86160\\
\hline
3&$2s$ & $2s$ & $2s$ & $2s$ & $4U_3$ & $0.74901$\\
3&$2s$ & $2s$ & $1d_{+}$ & $1d_{-}$ & $2U_3$ & $0.13949$\\
3&$2s$ & $1d_{+}$ & $2s$ & $1d_{-}$ & $2U_3$ & $0.66823$\\
3&$2s$ & $1d_{+}$ & $1d_{+}$ & $2s$ & $2U_3$ & $0.13949$\\
3&$2s$ & $1d_{-}$ & $2s$ & $1d_{-}$ & $2U_3$ & $0.66823$\\
3&$2s$ & $1d_{-}$ & $1d_{-}$ & $2s$ & $2U_3$  & $0.13949$\\
3&$1d_{+}$ & $1d_{+}$ & $1d_{+}$ & $1d_{+}$ & $3U_3$ & $0.71595$\\
3&$1d_{+}$ & $1d_{-}$ & $1d_{+}$ & $1d_{-}$ & $3U_3$ & $0.71595$\\
3&$1d_{+}$ & $1d_{-}$ & $1d_{-}$ & $1d_{+}$ & $3U_3$ & $0.12846$\\
3&$1d_{-}$ & $1d_{-}$ & $1d_{-}$ & $1d_{-}$ & $3U_3$ & $0.71595$\\
\hline 
\hline
\end{tabular}
\end{table}

For triangular and tetrahedral clusters of quantum dots we consider also
three-dimensional dots. This we do only within the $p$-shell where the
nonzero matrix elements are the same as in the two-dimensional case,
but now also including the $p_z$-orbitals.

\subsection{Hopping parameters}

In the simple Hubbard model with one state per site
the hopping is between two $s$-states and independent of the direction.
In the case of higher angular momentum states we have to consider also 
directional dependence. We have done this for the $p$-electrons
and $d$-electrons.

The (unnormalized) wave functions of the two-dimensional harmonic confinement
are
\begin{equation}
\label{trstates}
\begin{split}
\langle {\bf r}\vert 1s\rangle & = e^{- \alpha r^2} \\
\langle {\bf r}\vert p_\pm\rangle & =r e^{- \alpha r^2} e^{\pm i\phi}, 
           \qquad \left({\rm or}\quad\langle {\bf r}\vert p_\xi\rangle = \xi e^{- \alpha r^2}\right), \\
\langle {\bf r}\vert 2s\rangle & =(2 \alpha r^2 -1) e^{- \alpha r^2}, \\
\langle {\bf r}\vert d_\pm\rangle & =r^2 e^{- \alpha r^2} e^{\pm 2 i \phi}, \\
\end{split}
\end{equation}
where $\alpha$ depends on the oscillator parameter of the confinement, $\xi$ is either $x$ or $y$,
and $r=\sqrt{x^2+y^2}$. For $p$-electrons we use both angular momentum
presentation and Cartesian coordinates. Naturally the results are independent of
the choise of the coordinate system. We use the Cartesian coodinates
for studying three-dimensional dots with $p$-electrons. In that case 
$r=\sqrt{x^2+y^2+z^2}$ and $\xi$ can also be $z$.

\begin{figure}[h!!!]
\centering
\includegraphics[width=0.3\columnwidth]{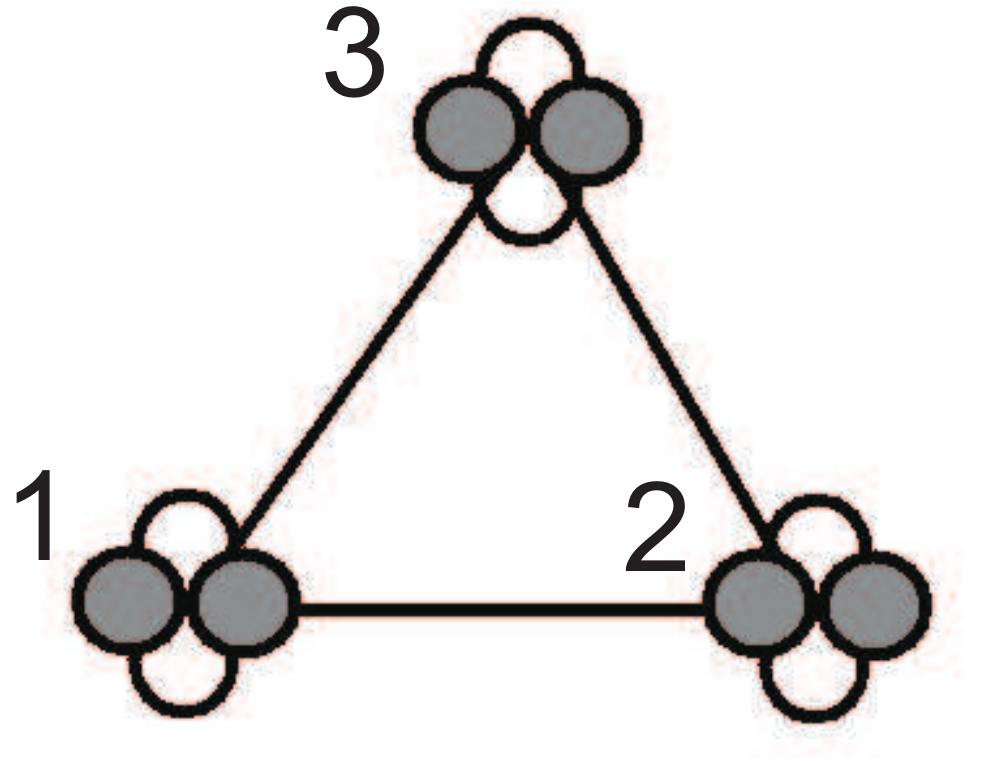}
\includegraphics[width=0.3\columnwidth]{nikkarilafig1a.pdf}
\includegraphics[width=0.3\columnwidth]{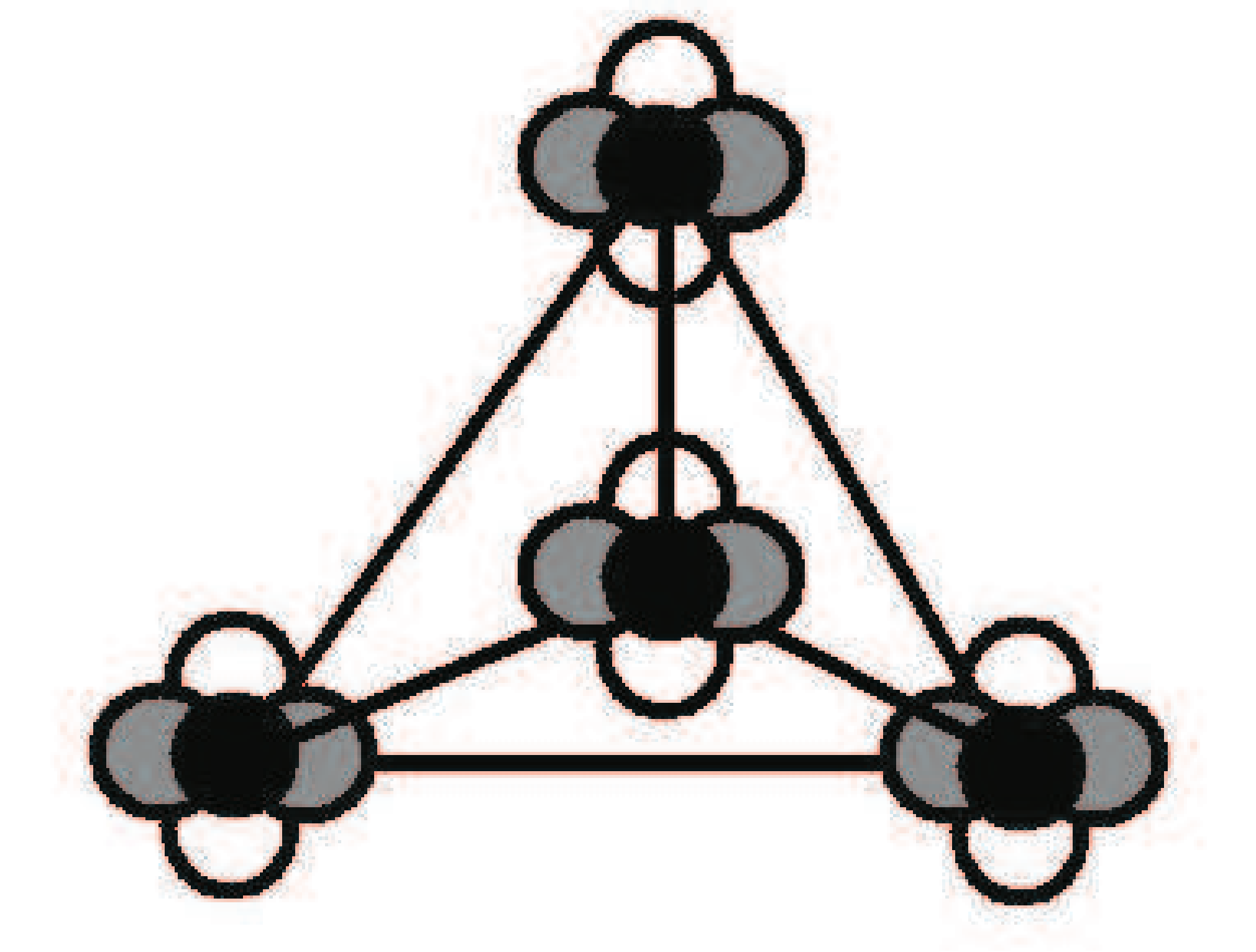} 
\caption{Schematic picture of triangular and tetrahedral clusters considered.
Left: triangle with $p_x$ and $p_y$ orbitals; center: triangle with $p_x$, $p_y$ and $p_z$ orbitals;
right: tetrahedron with $p_x$, $p_y$ and $p_z$ orbitals.}
\label{triangle}
\end{figure}

Figure \ref{triangle} shows triangular and tetrahedral clusters with $p$-orbitals.
At the vertices electrons can occupy states
$p_x$, $p_y$ (and also $p_z$ in 3D cases).
We assume that the hopping probability is proportional to the
overlap of the wave functions at neighboring sites.
The overlap integral depends on the factor $t=\alpha R^2$, where 
$R$ is the distance between the sites. Each nonzero
overlap is proportional to factor 
\begin{equation}
\label{constant}
A=\frac{32}{\alpha}e^{-\alpha R^2/2}= \frac{32}{\alpha}e^{-t/2}
\end{equation}
Some of the overlap integrals are zero due to the symmetry.
The overlap integrals of the $p_\xi$-electrons for a triangle and tetrahedron are given in
Table \ref{triangletab}. In the case of four dots in a square the hopping between
neighboring dots is along $x$ or $y$ direction and in this case the hopping 
from $p_x$ to $p_y$ is not possible. In the case of angular momentum
representation the hopping probability also depends on the direction due
the phases of the wave functions. Table \ref{jpsd} gives the hopping parameters
in $x$ and $y$ directions for the $p$ and $sd$ shells.
Note that the hopping probability is of the same order between all the states.

\begin{table}
\centering
\linespread{1}
\small
\caption{Hopping parameters $J_{nn'jj'}$ (in units of $A$) for $j$ and $j'$ 
orbitals ($p_x$, $p_y$ and $p_z$ orbitals)
at vertices $n$ and $n'$ of a triangle and a tetrahedron.
The symbol $n_1$ refers to triangle's vertices $1$ and $2$ (see Fig. \ref{triangle}),
whereas symbols $n_2$ and $n_3$ go through all $1$, $2$ and $3$ ($n_2 \neq n_3$).
The table for tetrahedron shows the additional hopping parameters from the
triangle to the fourth vertex. In the case of two signs, the upper sign is for 
$n_1=1$ and the lower sign for $n_1=2$.
}
\label{triangletab}

\begin{tabular}{|c|c|}
\hline
Triangle & Tetrahedron \\
\hline
\begin{tabular}{llllc}
$n$ & $n'$ & $j$ & $j'$ & $J_{nn'jj'}/A$ \\
\hline
$1$ & $2$     & $p_x$ & $p_x$ & $ ( 1-t )$  \\
$1$ & $2$     & $p_y$ & $p_y$ & $1$ \\
$1$ & $2$     & $p_x$ & $p_y$ & 0 \\
$n_1$ & $3$   & $p_x$ & $p_x$ & $ (1-t/4))$ \\
$n_1$ & $3$   & $p_y$ & $p_y$ & $(1-3 t/4)$ \\
$n_1$ & $3$   & $p_x$ & $p_y$ & $\mp t \sqrt{3}/4$\\
$n_2$ & $n_3$ & $p_x$ & $p_z$ & 0 \\
$n_2$ & $n_3$ & $p_y$ & $p_z$ & 0 \\
$n_2$ & $n_3$ & $p_z$ & $p_z$ & $ 1$ \\
\hline
 & & & & \\
  & & & &\\
 & & & & \\ 
\end{tabular} &
\begin{tabular}{llllc}
$n$ & $n'$ & $j$ & $j'$ & $J_{nn'jj'}/A$ \\
\hline
$n_1$ & $4$     & $p_x$ & $p_x$ & $ ( 1-t /4 )$  \\
$n_1$ & $4$     & $p_y$ & $p_y$ & $ ( 1-t /12 )$ \\
$n_1$ & $4$     & $p_x$ & $p_y$ & $\mp t/4  \sqrt{3}$ \\
$n_1$ & $4$     & $p_x$ & $p_z$ & $\mp t/ \sqrt{6}$ \\
$n_1$ & $4$     & $p_y$ & $p_z$ & $- t/3 \sqrt{2}$ \\
$n_1$ & $4$     & $p_z$ & $p_z$ & $( 1-2 t/3)$ \\
$3$ & $4$   & $p_x$ & $p_x$ & $1$ \\
$3$ & $4$   & $p_y$ & $p_y$ & $(1-t/3)$ \\
$3$ & $4$   & $p_x$ & $p_y$ & 0\\
$3$ & $4$     & $p_x$ & $p_z$ & 0 \\
$3$ & $4$     & $p_y$ & $p_z$ & $ \sqrt{2} t/3$ \\
$3$ & $4$     & $p_z$ & $p_z$ & $( 1-2 t /3)$ \\
\end{tabular}\\
\hline
\end{tabular}
\end{table}

We have only one parameter $t$
describing the hopping probability. Notice, however, that the
ratios of the different hopping parameters depend on $t$, i.e.
on the distance between the dots.
In real systems the parameter $t$ and the on-site 
interactions $U$ are coupled. However, we keep them as
independent parameters
in order to get a more general picture of the possible 
magnetic structures. In realistic systems where the tight-binding model
(and thus the Hubbard model) is expected to be valid, the
parameter $t$ should be of the order of 10 or larger. 
It turns out, however, that already when $t \gtrsim 4$ the results are
qualitatively independent of $t$.

\begin{table}
\caption{Hopping parameters for the $p$ shell and $sd$ shell calculated
for the angular momentum states. $J_{nn'jj'}^{\rm as}$ gives 
the asymptotic ratios ($t\rightarrow \infty$)
of the hopping parameters (note that even though $t\rightarrow \infty$,
$J_{nn'jj'}\rightarrow 0$ due to the prefactor $A$).
The hopping direction is along $x$ and $y$ axis. In the case of
two signs, the upper sign is for hopping to $x$-direction
and the lower sign to $y$-direction.}
\label{jpsd}
\begin{tabular}{|cc|c|c|}
\hline
$j$ & $j'$ & $J_{jj'}/A$ & $J_{nn'jj'}^{\rm as}/A$ \\
\hline
 $p_+$ & $p_+$ & $-(t-2)/2$ & $ -t/2$ \\
 $p_+$ & $p_-$ & $\mp t/2$ & $\mp t/2$ \\
\hline
$2s$ & $2s$ & $(-2+t)^2/4$ & $t$ \\
$2s$ & $d_+$ & $\pm t(-4+t)/4\sqrt{2}$ & $\pm t/\sqrt{2}$ \\
$d_+$ & $d_+$ & $(8-8t+t^2)/8$ & $t/2$ \\
$d_+$ & $d_-$ & $t^2/8$ & $t/2$ \\
\hline
\end{tabular}
\end{table}

\subsection{Heisenberg model}

It is well-known that in the limit of large $U/t$ the simple Hubbard
model with one $s$-state per site approaches the antiferromagnetic Heisenberg model. 
We will study if the clusters with half filled $p$ or $sd$ shells 
will also give the same spectrum as the Heisenberg model.
The effective Heisenberg Hamiltonian is
\be
\hat{H}_{\rm eff}=\frac{1}{2}J_{\rm eff}\sum_{n\ne n'}^L {\bf S}_n \cdot {\bf S}_{n'} 
+\text{constant}
\label{heisenberg},
\ee
where $L$ is the number of dots and
$J_{\rm eff}$ is the effective coupling which depends on the 
hopping parameters and the on-site interaction. The spin depends on the 
number of electrons in each dot and is determined by the Hund's first rule.
In the half-filled case the spin is 
1/2, 1, and 3/2 for the $1s$, $1p$ and $2s1d$ shells, respectively. 
Note that in most cases of the clusters considered here
the Heisenberg model is exactly solvable\cite{ashcroft1976,viefers2004}.

\section{Results}

\subsection{Quantum dot Dimer}

\begin{figure}[!h]
\includegraphics[width=0.98\columnwidth]{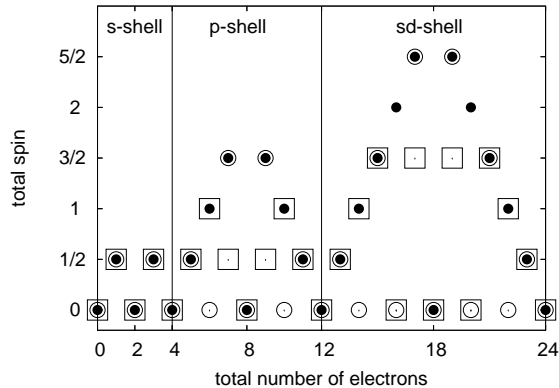} 
\caption{Total spin as a function of the number of electrons in 
a quantum dot dimer. The solid dots are the results for the
lateral and vertical dimer with 
delta function interaction. The 
open circles and squares are the results for Coulomb interaction
for lateral and vertiacal dimer, respectively.
In all cases the ratio of the largest $U$ and largens $J$ 
is $U_{\rm max}/J_{\rm max}=10$.
}
\label{dimer}
\end{figure}

We first consider two weakly connected quantum dots.
Figure \ref{dimer} shows the total spin as a function of
electrons in the quantum dot dimer. In the case of the full shells,
$N=4,\quad 12,\quad 24$,
the spin is naturally zero. In the case of half filled shells
the spin is also zero. However, in that case each dot has 
the spin determined by the Hund's rule, but the spins are opposite as in
an antiferromagnet. This was confirmed by looking at the 
conditional probability. 

Figure \ref{dimer} shows results for two lateral dots where the 
dots are in the same plane and for two vertical dots where the 
dots are on top of each other. In the lateral case the 
hopping probabilities are those given in Table \ref{jpsd} while 
in the vertical case the symmetry allows hopping is only between same orbitals
in both dots. In the case of the delta function interaction the
total spin is the same for lateral and vertical configurations.

In general, for an open shell quantum dot the Hund's rule determines
the spin\cite{reimann2002}. This is the case whatever is the the interactions
between the particles, as long as it is repulsive.
In the case of the delta function interaction the favored
magnetic coupling between the two dots is ferromagnetic. Only
the half-filled case is antiferromagnetic. 
In the case of the long-range Coulomb interaction the
situation is more complicated: There is a delicate
competition between the ferromagnetic and antiferromagnetic coupling,
and the total spin can also be between the largest possible and zero,
as seen in Fig. \ref{dimer}.  

The antiferromagnetic order of the half-filled shell can be 
seen also by looking at the energy spectrum of the lowest state.
The Heisenberg Hamiltonian, Eq. (\ref{heisenberg}), for two spins gives
energies $\epsilon(S)=S(S+1)J_{\rm eff}$, where $S$ is the spin of the state
which gets values $0,~ 1, \cdots S_{\rm max}$. Denoting by 
$K_S$ the ratio $K_S=(\epsilon(S)-\epsilon(0))/(\epsilon(1)-\epsilon(0))$ we 
notice that for the Heisenberg model $K_S=S(S+1)/2$. These ratios can now
be compared to those determined from energy spectrum of the Hubbard model.

\begin{table}[!h]
\caption{Ratios of energy differences $K_S$ of half-filled
double dots compared to those of the Heisenberg model.
The last column gives the ratio of the largest on-site energy to
the largest hopping integral.}
\label{kvalues}
\begin{tabular}{cccccrr}
Shell & System & Interaction& $K_1$ & $K_2$ &$K_3$ & $U_{\rm max}/J_{\rm max}$ \\
\hline
 & & Heisenberg & 1 & 3 & 6 \\
 \hline
$p$ & lateral & delta   & 1 & 3.06 & & 30.0 \\
$p$ & lateral & Coulomb & 1 & 3.20 & & 10.0 \\
$p$ & vertical   & delta   & 1 & 2.98 & & 10.0 \\
$p$ & vertical   & Coulomb & 1 & 2.68 & & 30.0 \\
\hline
$sd$ & lateral & delta   & 1 & 3.01 & 6.03 & 40.0\\
$sd$ & lateral & Coulomb & 1 & 3.01 & 6.04 & 75.0 \\
$sd$ & lateral & Coulomb & 1 & 3.44 & 11.93 & 7.5 \\
$sd$ & vertical   & delta   & 1 & 3.00 & 5.79 & 40.0 \\
$sd$ & vertical   & Coulomb & 1 & 2.93 & 5.71 & 15.0 \\
\hline
\end{tabular}
\end{table}

Table \ref{kvalues} shows the calculated energy ratios for 
different two dot molecules with half-filled $p$ or $sd$ shells.
The results show a good agreement with the antiferomagnetic Heisenberg model.
The results depend on the ratio of the on-site energies $U$ and
the hopping parameters $J$ as indicated in the case of lateral
Coulomb system. The larger is the ratio $U_{\rm max}/J_{\rm max}$
the better the Heisenberg model describes the lowest energy states.

In the case of the Coulomb interaction the spin is zero for all
even numbers of electrons. Each dot then has an integer number
of electrons and its spin is determined by the Hund's rule.
The conditional probability shows that the coupling between the
dots is antiferromagnetic. This result is in agreement with 
that of the local spin density approximation\cite{kolehmainen2000}.

\subsection{Triangle of quantum dots with with $p$ orbitals}

Next we study the magnetic properties of
trimers, where the quantum dots form vertices 
of an equilateral triangle as shown in Fig. \ref{triangle}.
We considered two cases: Two-dimensional quantum dots, where
only the $p_x$ and $p_y$ orbitals in each dot can be occupied 
and a three-dimensional case, where each dot is spherically
symmetric having also the $p_z$ orbital.
In the case of the triangle we only used the onsite
matrix elements calculated with the delta function interaction.

\begin{figure}[!h]
\includegraphics[width=0.98\columnwidth]{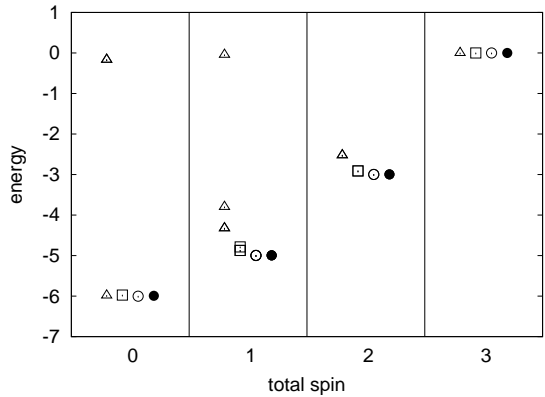} 
\caption{Lowest energy levels of the Hubbard model for a triangle
(without $p_z$ electrons) with half-filled $p$-shell compared with
the results of the antiferromagnetic Heisenberg model with spin $S=1$.
The distance between the dots $R$ is 3 (triangles), 3.5 (squares), 4 (circles)
and 5 (dots). In all cases $\alpha=1$ and $U_2=1$. The results for $R=5$ agree exactly
with those of the Heisenberg model. The energies have been scaled so that the
ground state energy is -6. 
}
\label{heis3xy}
\end{figure}

\begin{figure}[!h]
\includegraphics[width=0.98\columnwidth]{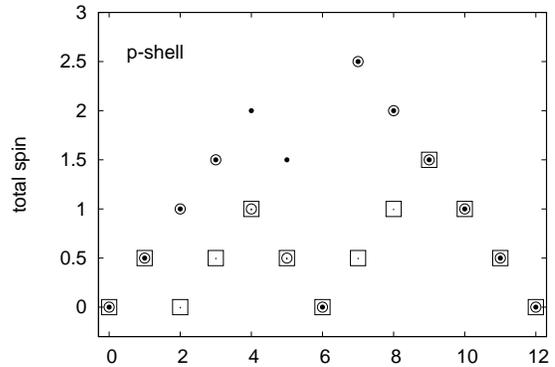}
\includegraphics[width=0.98\columnwidth]{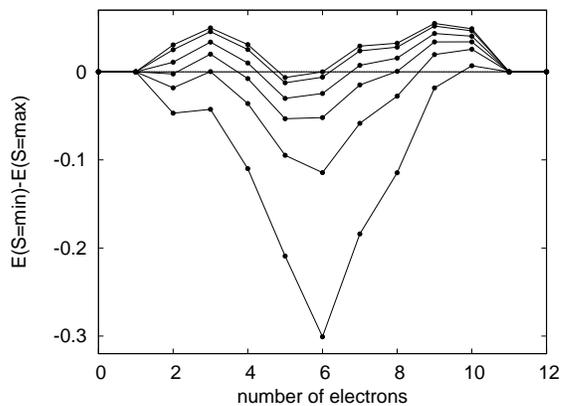}
\caption{Upper panel: Total spin as a function of the total 
number of $p$-electrons in a triangular cluster of quantum dots.
The interdot distance $R$ is 3 (squares), 3.5 (circles) and 4 (dots).
$U_2=1$. Lower panel: Energy difference between the $S=0$ (or $S=1/2$)
state and the polarized state with $S=N/2$ calculated for $U_2=1$ and
for different values of the hopping parameter $\gamma J$, where
$J$ is determined with $R=4$ and $\gamma$ has values 
0.01, 1, 4, 8, 16, 100 (from top to down). 
}
\label{trise}
\end{figure}

Figures \ref{heis3xy} and \ref{trise} show results of the
2D case. 
In the case of the half-filled orbital one would expect 
antiferromagnetic coupling. Indeed, the total spin 
of the ground state is zero for $N=6$. 
In the case of the triangle the antiferromagnet is frustrated.
The conditional probability is then not very helpful for confirming the
antiferromagnetism. However, in the case of a triangle
the antiferromagnetic Heisenberg Hamiltonian can be solved exactly
and the energy specrum can be compared to that of the Hubbard
model. This is done in Fig. \ref{heis3xy} for different values of
the interdot distance, i.e hopping probability. 
The figure shows that all the results are in 
agreement with those of the Heisenberg model, which becomes 
more accurate when the interdot distance increases. 
In the case of a triangle the Heisenberg model has only one energy state for each
total spin with degeneracies 1, 2, 3, and 1 for $S=0$, 1, 2, and 3, 
respectively.

Figure \ref{trise} shows the ground state spin as a function of the
total number of electrons in the triangle for three different
values of the hopping parameter. In the weak coupling limit (black dots)
the spin is at maximum for $N>6$, 
meaning ferromagnetic coupling between the dots.
For less than half-filled case, $N<6$, the situation is more complicated.
The weak coupling limit (black dots) show maximum spin, 
except for $N=5$ where $S=3/2$ (instead of the largest possible 5/2). 
When the coupling between the dots gets stronger, the total spin gets smaller and
the tendency to ferromagnetism disappears. This is illustrated also
in the energy plot in Fig. \ref{trise} where the energy difference between the
$S=0$ (or $S=1/2$) and the maximum spin state is plotted for different
values of the coupling. Weak coupling favors ferromagnetism except for the 
half-filled case while strong coupling favors small total spin.

We also computed the triangle of quantum dots including also the 
$p_z$ orbitals orthogonal to the plane of the triangle. 
Naturally, in this case each dot can occupy 6 electrons and a half-filled
dot will have spin 3/2 due to Hund's rule. The triangle with half-filling
($N=9$) had again low-energy spectrum in perfect agreement with the 
Heisenberg model for spin 3/2. For other fillings the ferromagnetic order
was favored in the weak coupling case, just like in the case of the triangle
without the $p_z$ orbitals. It is interesting to note that the 
antiferromagnetic Heisenberg model for half-filled case seems to be 
independent of how the coupling between the individual dots is made.
For example, when the $p_z$ orbitals are included in the weak coupling limit, 
the hopping between them is nearly prohibited and their role is just to
increase the spin per dot, i.e. the spin of the Heisenberg model.

\subsection{Tetrahedron with $p_x$ and $p_y$ and $p_z$ orbitals on 
its vertices}

The only three-dimensional cluster we considered is a tetrahedron.
In this case we only studied the $p$ shell
and included all the three $p$-orbitals. 
We considered only the delta function interaction and used
the Cartesian $p$-orbitals, the overlaps of which are given 
in Table \ref{triangletab}.

\begin{figure}[!h]
\includegraphics[width=0.98\columnwidth]{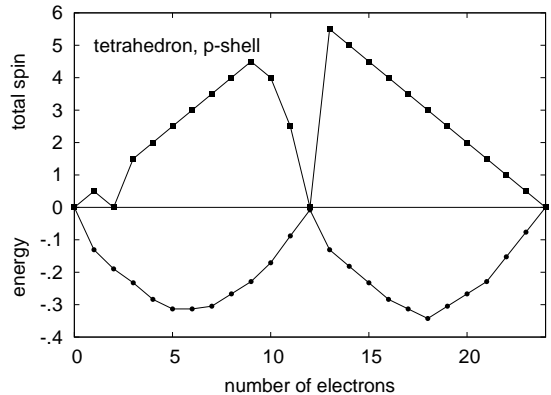} 
\caption{Total spin and ground state energy of four quantum dots
forming a tetrahedron, as a function of the total number of
$p$-electrons (including $p_z$ electrons). The distance between the dots 
is $R=4$ and the $U_2=1$. Note that the energy for the antiferromagnetic
ground state for $N=12$ is not zero but $E=-.007533$. 
}
\label{tetrah}
\end{figure}

The results show that for strong hopping (small distance between the
dots) and weak interaction
(small $U_2$) the total spin is mainly 0 or 1 indicating no magnetism.
For large $U_2$ or weak coupling between the dots
the results are closely related to those observed for a triangle.
Figure \ref{tetrah} shows the total ground state energy and 
the total spin as a function of the number of electrons.
The parameters correspond to the weak coupling limit.
Like in the case of a triangle, the system is ferromagnetic
for more than half-filling of the $p$ shell ($N>12$),
while for less than half filling the spin is slighly reduced for 
some electron number. In exactly half-filled case the result is
again a Heisenberg antiferromagnet with a very good accuracy.

The total energy plotted in Fig. \ref{tetrah} shows that it has 
minima at $N=5$ and $N=18$ corresponding roughly 1/4 and 3/4 fillings.
Note that for $N>12$ the energy shown is the ground state energy from 
which the average on-site repulsion, $U_2(N-12)$, has been substracted.
In the half-filled case the energy is only slightly negative, arising
from 'virtual' hopping between dots. 

The asymmetry of the total spin (and energy) with respect of the
half filling in the case of triangle and tetrahedron can be traced 
back to the single particle levels of the tight binding model in these
systems: They are not symmetrically distributed around the zero energy.

\subsection{Four dots in a square and in a row: The $p$-shell}

\begin{figure}[!h]
\includegraphics[width=0.98\columnwidth]{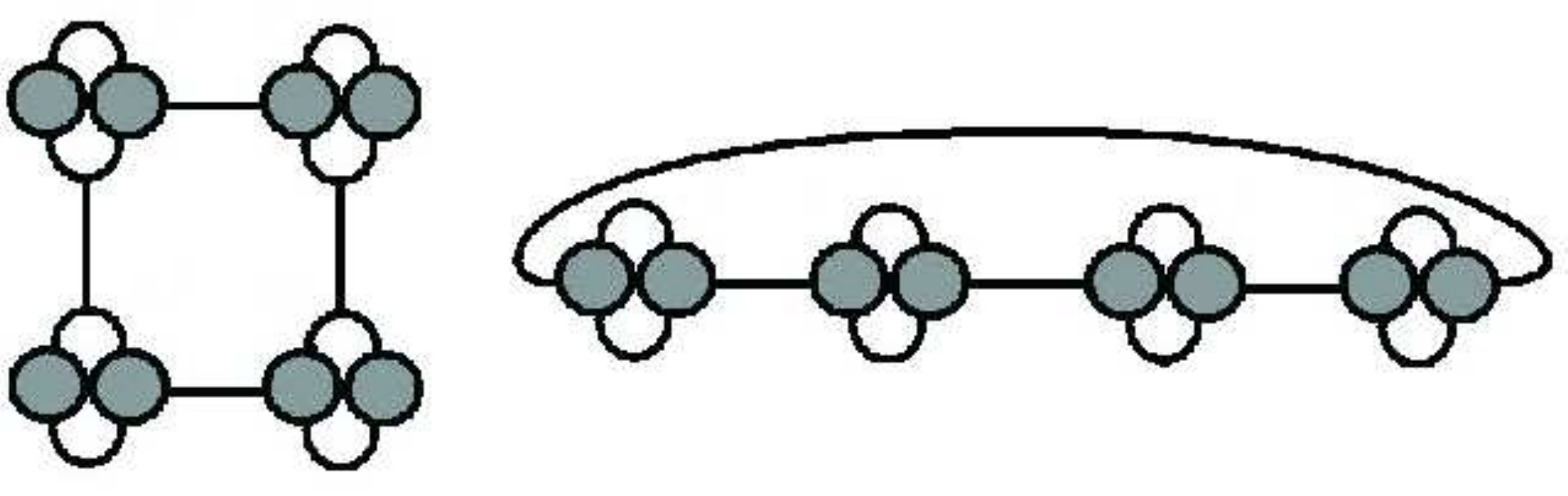} 
\caption{Schematic picture illustrating the easy hops between 
$p$electrons in a square and in a row with periodic boundary condition.
}
\label{squares}
\end{figure}

In the case of a square the hopping parameters are easier
to determine. In the case of $p$ orbitals we have
only hopping from $p_x$ to $p_x$ and from $p_y$ to $p_y$
as shown in the first two rows of the Table \ref{triangletab}
for triangles. The hopping is more favorable when the orbitals
point towards the neighboring dot. In the square this leads 
to an interesting situation where an electron, say in $p_x$ orbital,
can easily hop on the $x$-direction, but can not continue
easily around the square since every second hop would be in the
$y$-direction. The square is thus different from a row of four 
dots with periodic boundary conditions, as illustrated
in Fig. \ref {squares}. In the latter case the electron can continue
through the row with easy hops. 

\begin{table}
\caption{Total spin of the ground state for
a four dot square and row and row with periodic
boundary conditions (noted a Row+). 
$N$ is the total number of $p$ electrons in the four
dot system. C refers to Coulomb interaction and $\delta$
to delta function interaction. $U_{\rm max}/J_{\rm max}=17.2$ for
Coulomb interaction and 20 for delta function interaction.
In the cases of square with 2 and 14 electrons the ground state is
degenerate with spins 0 and 1.}
\label{squarep}
\begin{tabular}{c|c|c|c|c}
\hline
$N$ & Square C & Square $\delta$ & Row $\delta$ & Row+ $\delta$\\
\hline
2 & 0,1 & 0,1 & 0 & 1 \\
 4 & 0 & 2 & 2 & 0 \\
 6 & 0 & 0 & 3 & 3 \\
8 & 0 & 0 & 0 & 0 \\
10 & 0 & 0 & 3 & 3 \\
12& 0 & 2 & 2 & 0 \\
14& 0,1 &0,1 & 0 & 1 \\
\hline
\end{tabular}
\end{table}

Table \ref{squarep} shows the total spin of the ground state of the 
four dot square as a function of the filling of the $p$ shell.
The result is compared to that of the row of four dots and the row 
with periodic boundary conditions. 
In the case of the Coulomb interaction the total spin of the
of the ground state of the square of quantum dots
is zero for all (even) numbers of $p$-electrons. In the case of
the delta function interaction the cases with $N=4$ and $N=12$
favor maximizing the spin. The row of four dots favors ferromagnetism.
It is interesting to note that the difference in the hopping probabilities
between $p_x$ and $p_y$ orbitals seem to be important for the magnetic 
coupling between the dots. For example, for $N=6$ and $N=10$ the 
total spin is zero in the case of the square, while it has its maximum
value in the case of a row with periodic boundary conditions.

\begin{figure}[!h]
\includegraphics[width=0.98\columnwidth]{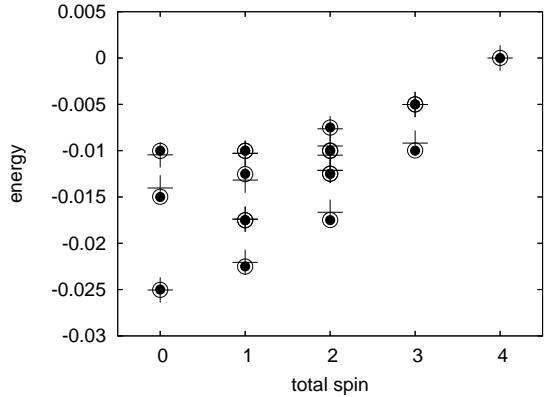} 
\caption{Energy spectrum of a square of quantum dots with 
half-filled $p$-levels (N=8). Black dots: results of the
antiferromagnetic Heisenberg model. Open circles: results
of the Hubbard model with delta function interaction and
$U_{\rm max}/J_{\rm max}=40$. Plus-signs: results of the 
Hubbard model with Coulomb interaction and $U_{\rm max}/J_{\rm max}=17.2$.
The energies of the Heisenberg model and Coulomb interaction are scaled and 
shifted so that the lowest and highest energies shown agree with those of the
Hubbard model with the delta function interaction.}
\label{sqpheis}
\end{figure}

In the half-filled case each system has zero total spin.
Figure \ref{sqpheis} shows the comparison of the lowest energy
states of the half-filled case with those of the Heisenberg
antiferromagnetic square with $S=1$. For the delta function
interaction the agreement is perfect. In the case of the long-range
Coulomb interaction the quantitave agreement is not exact
but qualitatively the spectra are still the same. 
It is important to note that the directional dependence of the hopping,
i.e. the fact that the electron can not go easily around the square without
exchanging the orbital from $p_x$ to $p_y$, does not seem to have any effect
on the validity of the Heisenberg model.

\subsection{Four dots in a square and in a row: The $sd$-shell}

In this section we consider four dot systems
where in each dot the $sd$ shell is partially filled. 
We assume the dots to be two-dimensional. The open shell
then has $2s$, $d_+$ and $d_-$ orbitals.
The on-site interactions are given in Table \ref{umatrix}
and the hopping integrals in Table \ref{jpsd} above. 
For simplicity we use only the asymptotic values 
of the hopping parameters and, consequently, we have
only one free parameter which is the ratio 
$U_{\rm max}/J_{\rm max}$.

\begin{figure}[!h]
\includegraphics[width=0.98\columnwidth]{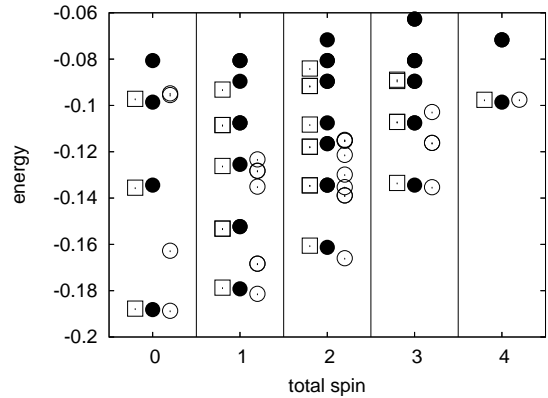} 
\caption{25 lowest energy states for 12 $sd$ electrons in a square
calculated with the delta function ($U_{\rm max}/J_{\rm max}=10$), squares, and with the 
Coulomb interaction ($U_{\rm max}/J_{\rm max}=7.5$), open circles,
compared with the results of the Heisenberg model, black dots. 
The results for the Coulomb 
interaction and the Heisenberg model have been scaled and shifted 
so that the ground state and the lowest
$S=4$ state agree with those of the delta function interaction.
}
\label{sqsdheis}
\end{figure}

\begin{figure}[!h]
\includegraphics[width=0.98\columnwidth]{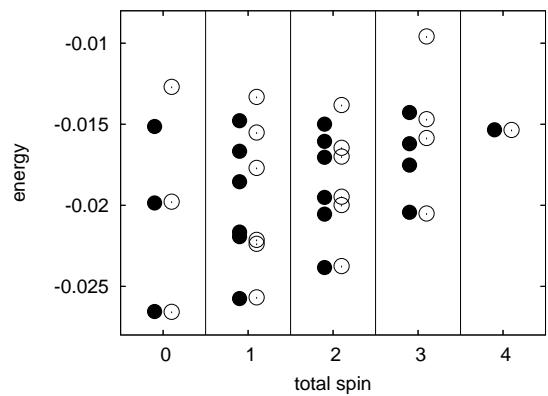} 
\caption{25 lowest energy states for 12 $sd$ electrons in a row of four
quantum dots. Black points are results for a row of vertical dots
calculated with the delta function interaction  ($U_{\rm max}/J_{\rm max}=40$),
open circles are results for a row of lateral dots
calculated with the 
Coulomb interaction ($U_{\rm max}/J_{\rm max}=7.5$). 
The results Coulomb interaction have been scaled and shifted 
so that the ground state and the lowest
$S=4$ state agree with those of the delta function interaction.
}
\label{rowsdheis}
\end{figure}

We first studied the ground state spin as a function of
the total (even) number of electrons for a square of quantum
dots using the Coulomb interaction. 
In the case $N=2$ and $N=22$ electrons the ground state is
degenerate with spins 0 and 1. For $N=4$ the total spin of the ground state
is 2 and for $N=10$ it is 1. For all other even numbers of electrons
the spin of the ground state is zero. 
In the case of a row of four lateral dots we studied the ground state only
up to half filling. The total spin for the row is zero for $N=2,~ 4,~ 6, ~ 8$ and 12,
but for $N=10$ it is 3.
Kolehmainen {\it et al}\cite{kolehmainen2000} studied the square of 
four quantum dots with 16 $sd$-electrons using the local spin density approximation.
They found the ground state to be ferromagnetic which is in disagreement with
the present result $S=0$, suggesting that the LSDA can not describe
correctly the strongly correlated system in this case.

In the case of half-filling, i.e. with $N=12$, we compared the low
energy spectra with that of the Heisenberg model with $S=3/2$.
Figure \ref{sqsdheis} shows the 25 lowest energy states for 
the square of four dots calculated with the Coulomb and 
delta function interactions,  compared to the results of the 
Heisenberg model. The delta function interaction reproduces 
nearly perfectly the spectrum of the Heisenberg model. 
Note that more levels are shown for the Heisenberg model and
that some of the levels are degenerate.
In the spectrum of the Coulomb interaction the lowest state for
each spin is still in fair agreement with the Heisenberg model,
but the higher levels do not agree.

We also studied in detail the spectrum of half-filled row of four dots.
In this case we dot not have an analytic solution for the Heisenberg model.
Instead we computed the row of four dots in the vertical arrangement
using the delta function interaction and using a large ratio
$U_{\rm max}/J_{\rm max}=40$, which we believe gives nearly exactly
the result of the Heisenberg model (for $S=3/2$). Figure \ref{rowsdheis}
shows a comparison of that result with the spectrum of
a lateral row of four dots with Coulomb interaction.
In this case the agreement of with the stecra is surprisingly good.

\section{Conclusions}

We have studied the magnetism of the ground sates of
clusters of quantum dots. 
We described each quantum dot with a harmonic confinement
and used a generalized Hubbard model for describing the interaction 
between the dots. For the on-site interaction we used
either the Coulomb interaction or the delta function interaction.
Only the partially filled harmonic oscillator shell was
used as an active basis of the many-particle calculation.

The quantum dot dimer favors ferromagnetic ground state
when the interaction is short range delta function. Only
the half-filled shells have antiferromagnetic order and
energy spectrum in good agreement with the Heisenberg model.
The case of Coulomb interaction is more complicated, but 
also in this case the half-filled shells agree with the 
Heisenberg model.

An equilateral triangle of three dots and a tetrahedron
of four dots was studied in the case of partially filled $p$-shell. 
Also in these frustrated cases the half-filled cases 
are well described with the antiferromagnetic Heisenberg model
while otherwise ferromagnetic coupling between the dots is favored.

In the case of lateral four dot molecules we studied the row and the square.
The half-filled case is again described well with the Heisenberg
model. When the interaction is the delta function repulsion
the agreement is nearly perfect and for Coulomb interaction
also surprisingly good. This is true for both the $p$-shell 
and the $sd$-shell. 
For other fillings the Coulomb interaction seems to favor
small or zero spin for the ground state.

{\bf Acknowledgments}

This work was supported by the Academy of Finland and by the 
Jenny and Antti Wihuri Foundation.

\end{document}